\font\bbb=msbm10                                                   

\def\R{\hbox{\bbb R}}

\def\ACMCS{{\sl ACM Comput.\ Surveys}}
\def\AcM{{\sl Acta Math.}}
\def\AI{{\sl Artif.\ Intell.}}

\def\ANASP{{\sl Ann.\ Scuola Norm.\ Sup.\ Pisa}}
\def\ARFM{{\sl Ann.\ Rev.\ Fluid Mech.}}

\def\BSTJ{{\sl Bell System Tech.\ J.}}

\def\CE{{\sl Comput.\ Econom.}}
\def\CDE{{\sl Contrib.\ Diff.\ Eq.}}

\def\E{{\sl Econometrica}}
 
\def\ESQ{{\sl Econom.\ Studies Quart.}}

\def\IER{{\sl Int.\ Econom.\ Rev.}}

\def\JCP{{\sl J. Comput.\ Phys.}}

\def\JET{{\sl J. Econom.\ Theory}}
\def\JFM{{\sl J. Fluid Mech.}}

\def\JME{{\sl J. Math.\ Econom.}}

\def\JMPA{{\sl J. Math.\ Pures Appl.}}

\def\JRAM{{\sl J. Reine Angew.\ Math.}}
\def\JSP{{\sl J. Stat.\ Phys.}}

\def\MA{{\sl Math.\ Ann.}}

\def\MMP{{\sl Monat.\ Math.\ Phys.}}

\def\Na{{\sl Nature}}

\def\PRSE{{\sl Proc.\ Roy.\ Soc.\ Edinburgh}}
\def\PRSLA{{\sl Proc.\ Roy.\ Soc.\ Lond.\ A}}

\def\RAS{{\sl Robotics and Autonomous Systems}}

\def\SIAMJNA{{\sl SIAM J. Numer.\ Anal.}}

\def\SRO{{\sl Sociological Research Online}}
\def\TAMS{{\sl Trans.\ Amer.\ Math.\ Soc.}}

\def\dajm{\hbox{D. A. Meyer}}

\def\brosl{\hbox{B. Hasslacher}}

\def\Segel{1}
\def\rMB{2}
\def\rBH{3}
\def\Sorkin{4}
\def\Kelvin{5}
\def\Helmholtz{6}
\def\Leonard{7}
\def\Ruijgrok{8}
\def\rSMD{9}
\def\Arrow{10}
\def\rBP{11}
\def\Rosenhead{12}
\def\Aref{13}
\def\Christiansen{14}
\def\Chorin{15}
\def\Hald{16}
\def\rSL{17}
\def\Scarf{18}
\def\Walras{19}
\def\Brouwer{20}
\def\Uzawa{21}
\def\rAD{22}
\def\Condorcet{23}
\def\rPB{24}
\def\rKS{25}
\def\rSPAKM{26}
\def\rSB{27}
\def\rMBLAEA{28}
\def\rZRER{29}
\def\rWCW{30}
\def\rHT{31}
\def\rBWM{32}
\def\rHM{33}
\def\rGDT{34}
\def\Godel{35}
\def\rVD{36}

\def\hfb{\hfil\break}

\catcode`@=11
\newskip\ttglue

   \font\ninerm=cmr9    \font\eightrm=cmr8   \font\sixrm=cmr6
  \font\ninebf=cmbx9   \font\eightbf=cmbx8  \font\sixbf=cmbx6
  \font\nineit=cmti9   \font\eightit=cmti8  
  \font\ninesl=cmsl9   \font\eightsl=cmsl8  
  \font\ninemi=cmmi9   \font\eightmi=cmmi8  \font\sixmi=cmmi6

\font\bigten=cmr10 scaled\magstep2 

\def\ninepoint{\def\rm{\fam0\ninerm}%
  \textfont0=\ninerm \scriptfont0=\sixrm
  \textfont1=\ninemi \scriptfont1=\sixmi
  \textfont\itfam=\nineit  \def\it{\fam\itfam\nineit}%
  \textfont\slfam=\ninesl  \def\sl{\fam\slfam\ninesl}%
  \textfont\bffam=\ninebf  \scriptfont\bffam=\sixbf
    \def\bf{\fam\bffam\ninebf}%
  \tt \ttglue=.5em plus.25em minus.15em
  \normalbaselineskip=11pt
  \setbox\strutbox=\hbox{\vrule height8pt depth3pt width0pt}%
  \normalbaselines\rm}

\def\eightpoint{\def\rm{\fam0\eightrm}%
  \textfont0=\eightrm \scriptfont0=\sixrm
  \textfont1=\eightmi \scriptfont1=\sixmi
  \textfont\itfam=\eightit  \def\it{\fam\itfam\eightit}%
  \textfont\slfam=\eightsl  \def\sl{\fam\slfam\eightsl}%
  \textfont\bffam=\eightbf  \scriptfont\bffam=\sixbf
    \def\bf{\fam\bffam\eightbf}%
  \tt \ttglue=.5em plus.25em minus.15em
  \normalbaselineskip=9pt
  \setbox\strutbox=\hbox{\vrule height7pt depth2pt width0pt}%
  \normalbaselines\rm}

\def\sfootnote#1{\edef\@sf{\spacefactor\the\spacefactor}#1\@sf
      \insert\footins\bgroup\eightpoint
      \interlinepenalty100 \let\par=\endgraf
        \leftskip=0pt \rightskip=0pt
        \splittopskip=10pt plus 1pt minus 1pt \floatingpenalty=20000
        \parskip=0pt\smallskip\item{#1}\bgroup\strut\aftergroup\@foot\let\next}
\skip\footins=12pt plus 2pt minus 2pt
\dimen\footins=30pc

\def\ie{{\it i.e.}}

\def\@versim#1#2{\lower.5pt\vbox{\baselineskip0pt \lineskip-.5pt
    \ialign{$\m@th#1\hfil##\hfil$\crcr#2\crcr\sim\crcr}}}
\def\gsim{\mathrel{\mathpalette\@versim\>}}

\magnification=1200
\input epsf.tex

\dimen0=\hsize \divide\dimen0 by 13 \dimendef\chasm=0
\dimen1=\hsize \advance\dimen1 by -\chasm \dimendef\usewidth=1
\dimen2=\usewidth \divide\dimen2 by 2 \dimendef\halfwidth=2
\dimen3=\usewidth \divide\dimen3 by 3 \dimendef\thirdwidth=3
\dimen4=\hsize \advance\dimen4 by -\halfwidth \dimendef\secondstart=4
\dimen5=\halfwidth \advance\dimen5 by -10pt \dimendef\indenthalfwidth=5
\dimen6=\thirdwidth \multiply\dimen6 by 2 \dimendef\twothirdswidth=6
\dimen7=\twothirdswidth \divide\dimen7 by 4 \dimendef\qttw=7
\dimen8=\qttw \divide\dimen8 by 4 \dimendef\qqttw=8
\dimen9=\qqttw \divide\dimen9 by 4 \dimendef\qqqttw=9

\parskip=0pt\parindent=0pt

\line{\hfil October 1997}
\line{\hfil chao-dyn/9710005}
\bigskip\bigskip\bigskip
\centerline{\bf\bigten TOWARDS THE GLOBAL:}
\medskip
\centerline{\bf\bigten COMPLEXITY, TOPOLOGY AND CHAOS}
\medskip
\centerline{\bf\bigten IN MODELLING, SIMULATION AND COMPUTATION}
\vfill
\centerline{\bf David A. Meyer}
\bigskip
\centerline{\sl Center for Social Computation,
                Institute for Physical Sciences, and}
\centerline{\sl Project in Geometry and Physics}
\centerline{\sl Department of Mathematics}
\centerline{\sl University of California/San Diego}
\centerline{\sl La Jolla, CA 92093-0112}
\centerline{dmeyer@chonji.ucsd.edu}
\vfill
\centerline{ABSTRACT}
\bigskip
\noindent Topological effects produce chaos in multiagent simulation 
and distributed computation.  We explain this result by developing 
three themes concerning complex systems in the natural and social 
sciences:  ({\it i\/})  Pragmatically, a system is complex when it is 
represented {\sl efficiently\/} by different models at different 
scales.  ({\it ii\/})  Nontrivial topology, identifiable as we scale 
towards the global, {\sl induces\/} complexity in this sense.  
({\it iii\/})  Complex systems with nontrivial topology are typically 
chaotic.
\bigskip\bigskip\bigskip
\global\setbox1=\hbox{Key Words:\enspace}
\parindent=\wd1
\noindent{\sl Journal of Economic Literature\/} Classification System:
                   D71,      
                   D72,      
                   D50,      
                   C62.      

\noindent Physics and Astronomy Classification Scheme:
                   05.45.+b, 
                   47.11.+j, 
                   95.10.Ce. 

\noindent American Mathematical Society Subject Classification:
                   54H20,    
                   58F10,    
                   58F21.    

\item{Key Words:}  complex systems, scale hierarchy, multiagent 
                   simulations, Condorcet cycle, Arrow's theorem, 
                   excess demand theorems, chaotic dynamics.

\vfill
\hrule width2.0truein
\medskip
\noindent Expanded version of a talk presented at the International 
Conference on Complex Systems held in Nashua, NH, 21--26 September 
1997.
\eject

\headline{\ninepoint\it Towards the global       \hfil David A. Meyer}
\parskip=10pt
\parindent=20pt

\noindent{\bf 1.  Introduction}

\noindent Although the concerns of modelling and simulation can be 
quite different [\Segel], both support a pragmatic definition of 
complexity:  {\sl A system is complex if it is represented efficiently 
by different models at different scales.}  This idea is commonplace,
reflected in the way we organize our understanding of the world around 
us into physics, chemistry, biology, psychology, economics and 
political science at (roughly) increasing size scales.  The goal of 
this paper is to explain a recently demonstrated difficulty with 
multiagent simulations of complex systems at the social science end of 
this spectrum [\rMB] by placing it in the context of (possibly more 
familiar) models and simulations of complex systems at the natural 
science end.

We begin by showing that this definition of complexity is more than 
subjective.  In Section~2 we consider the hierarchical algorithm of 
Barnes and Hut for simulation of gravitationally interacting particles 
[\rBH].  Their multiscale algorithm is more efficient, in a very 
precise sense, than na{\"\i}ve direct simulation.  Large scale states
are described by total mass and average position of particles, so the
model is similar at different scales.  We describe this situation as
{\sl simplicity}.  For more complicated systems, as we increase in 
scale towards the global, we may identify states with nontrivial 
topology.  The presence of such states can lead to a different and 
more efficient model at the larger scale; {\sl topology induces 
complexity}.  

The purest example of this phenomenon might be topological geons in 
quantum gravity [\Sorkin]:  prime components of the spacelike 
3-manifold comprise localized nontrivial topology and can be modelled 
as particles.  The intellectual heritage of this model includes Lord 
Kelvin's theory of atoms as knotted vortex lines [\Kelvin].  That 
theory was mistaken, of course, but it was motivated by Helmholtz's 
mathematical analysis of hydrodynamical vortices [\Helmholtz].  
Vortices alone {\sl are\/} topologically nontrivial states, in terms 
of which the equations for fluid mechanics can be recast and which are 
utilized in very practical vortex simulations of fluid flow 
[\Leonard].  We explain this in Section~3, remarking on the 
efficiencies gained by use of a hierarchical algorithm.  While 
demonstrations of improved simulation efficiency are not available for 
all the systems we consider, these two examples reenforce the belief, 
based on the conceptual adequacy of different models at different 
scales, that the systems are indeed complex.

Multiagent systems are discrete, in contrast to PDE models for fluid
mechanics.  To develop our theme of topology induced complexity for 
application in the former, we must explain how it can occur in 
non-continuum systems.  In Section~4 we consider the homogeneous 
sector of a multispecies reaction-diffusion model in 
chemistry/population biology.  This model, analyzed by Ruijgrok and 
Ruijgrok [\Ruijgrok], illustrates our first two themes:  (discrete) 
topology induces complexity in the sense of a different efficient 
model at the global scale.

The same nontrivial discrete topology---a cycle---is ubiquitous in 
formal models of economics and political science:  In Sections~5 and 6 
we describe two fundamental results---Sonnenschein, Debreu, and 
Mantel's excess demand theorems [\rSMD] and Arrow's voting theorem 
[\Arrow]---which guarantee the existence of cycles in market and 
voting models, respectively.  We also expand our third theme:  systems 
with topologically induced complexity are typically chaotic.  In 
particular, we explain the precise mathematical sense in which 
aggregation by voting makes multiagent simulations chaotic [\rMB].

But the aggregation processes which scale multiagent simulations 
towards the global are usually market mechanisms or voting rules; 
thus such systems are typically complex, and, according to the results 
described in Section~6, chaotic.  In the final section we discuss the 
consequences of this phenomenon and mechanisms which might be 
implemented to control it.

\medskip
\noindent{\bf 2.  Hierarchical efficiency}

\noindent Simulation of interesting natural or social systems compels 
careful attention to the efficiency of the algorithms used.  
Algorithms which are exponential in the size of the system are 
essentially useless, and even for polynomial algorithms, decreasing 
the leading exponent reduces runtime by orders of magnitude.  The 
simulations of concern here typically consist of a large number $N$ of 
fundamental objects interacting according to rules which model the 
dynamics of the system under investigation.  Algorithmic efficiency is 
thus determined by runtime as a function of $N$.

Consider the problem of simulating the (Newtonian) gravitational 
dynamics of $N$ particles.  Each particle exerts a force on every 
other particle and is subject to the sum of the forces exerted on it 
by all the other particles.  The most straightforward algorithm would 
compute the $N(N-1)/2$ pairwise forces, sum the forces on each 
particle, and then evolve each particle accordingly, at each timestep.
This algorithm involves no approximations beyond the finite precision
computer representation of real numbers; it provides an accurate 
$O(N^2)$ description of the dynamics.

Barnes and Hut showed, however, that by aggregating the particles into
a hierarchy of clusters of increasing sizes, the average run time can 
be reduced to $O(N\log N)$ {\sl with bounded error\/} [\rBH].  Their
algorithm works by dividing up the volume of space containing the 
particles into a tree of cubical cells:  Starting with a cube large 
enough to contain all the particles, at each timestep consider the 
particles in some (arbitrary) sequence.  While the particle lies in 
the same cube as any previous particle, subdivide that cube into eight
cubes of half the linear size.  Now assign to each non-empty cube a
`cluster-particle' with the total mass of the particles in that cell, 
and locate the cluster-particle at the center of mass of those 
particles.  On average, constructing the tree of cells and
cluster-particles requires $O(N\log N)$ steps.

The force on a given particle is now computed recursively by working
down the tree:  If a cluster-particle lies in a cube of size $l$, is 
distance $d$ from the particle, and $l/d < \theta$ (a constant), 
compute the force it would exert; if not move down to the next smaller
cluster-particles in the tree and repeat.  This algorithm approximates 
the force on a particle, with bounded error (depending on $\theta$), 
in an average of $O(\log N)$ steps.  The average runtime per timestep 
is thus $O(N\log N)$ for the whole algorithm.

In this description `average' refers to possible particle 
configurations with respect to a uniform probability distribution on
the cube.  When the particles are literally clustered, as in the 
simulation Barnes and Hut present of two interacting clusters, 
additional efficiencies obtain.  Even in this case, however, the
larger scale representation is essentially similar to the smaller 
scale one; the system is simple.  

\medskip
\noindent{\bf 3.  Topology induces complexity}

\noindent Fluid flow provides our first example of a system with 
different representations at different scales.  Disregarding the fact
that real fluids are composed of molecules, which are composed of 
atoms, $\ldots$, and should, by virtue of that multiscale description
alone, be complex systems, let us take the system to be defined by the
Navier-Stokes equations:
$$
\eqalign{
\partial_t v +  v\cdot\nabla v &= -\nabla p + \nu\nabla^2 v        \cr
                 \nabla\cdot v &= 0,                               \cr
}
$$
together with appropriate boundary conditions.  Here $v$ is the 
velocity field of the (incompressible) fluid, $p$ is the pressure, and 
$\nu$ is the viscosity.  Immense effort has, of course, gone into
simulating the Navier-Stokes equations [\rBP], whose solutions range 
from laminar to turbulent flow.

The configuration variable in this system, the velocity field, can be 
topologically nontrivial:  there may be, for example, {\sl vortices}, 
\ie, closed loops in the flow.  Their presence is measured by the 
{\sl vorticity\/} $\omega := \nabla\wedge v$ [\Helmholtz], in terms of 
which the first equation above can be rewritten as
$$
\partial_t\omega = \nabla\wedge(v\wedge\omega) + \nu\nabla^2\omega,
$$
eliminating the pressure.  The inverse of the Poisson equation 
defining vorticity is the Biot-Savart law; it gives $v$ as a function 
of $\omega$ and allows the system to be reformulated entirely in terms 
of the vorticity.  This is a different, although entirely equivalent, 
model for fluid flow.

When the flow is restricted to two dimensions the vorticity has only a
single (orthogonal) component so the configuration variable for the
model is a (pseudo-)scalar field.  As such it can be approximated by
a superposition of delta functions, or more physically, by a 
collection of point vortices with circulations $c_i$:
$$
\omega(r,t) \approx \sum_{i=1}^N c_i \delta\bigl(r - r_i(t)\bigr).
$$
Even before the advent of digital computers Rosenhead simulated fluid 
flow using this approximation [\Rosenhead]:  For inviscid flow, the 
velocity of each vortex is the value of the velocity field at its 
present location; this is given in terms of the vorticity field by the
Biot-Savart law.  Notice that a straightforward algorithm for this 
computation would be $O(N^2)$.  Just as in the simulation of 
gravitating particles considered in Section~2, the vortex positions 
can then be integrated forward in time.

There are several problems with this procedure:  While it is correct 
for inviscid flow, the singularities in the vorticity field make the 
errors in time integration difficult to control when the point 
vortices pass too close to one another.  (We anticipate the results of
Section~6 by remarking that such systems with only a few vortices have 
been shown to be chaotic [\Aref].)  This problem can be resolved to 
some extent by implementing a `cloud-in-cell' technique 
[\Christiansen] which also improves the runtime for solving the 
Poisson equation to $O(N) + O(M\log M)$, where $M$ is the size of an 
approximating spatial mesh.  Alternatively, and at the same time 
modifying the model so as to be able to simulate viscous flow, the 
point vortices can be replaced with `vortex blobs' in which the 
vorticity has a fixed but nonsingular distribution [\Chorin].  Hald 
has shown that the error can be controlled in simulations of inviscid 
flow with such models [\Hald].

These simulations demonstrate that two dimensional hydrodynamics is
therefore a complex system according to our operational definition:  
Not only are there efficient algorithms at different size scales, as 
presaged by our discussion of gravitating particles, but the system is 
modelled differently at each scale---by a velocity field, by a 
collection of vortices, and by a distribution of vorticity on a mesh.  
Topology will play this role---inducing new models at more global 
scales---for the rest of the discussion.

\medskip
\noindent{\bf 4.  Finite topology}

\noindent Vortices in fluids contain cycles---closed continuous flow 
lines.  But continuity of this sort is not necessary for nontrivial
topology.  Moving away from physics to a model for a system in 
chemistry or population biology, consider three species (types of 
particles) with interactions%
\sfootnote*{Delightfully, Sinervo and Lively have observed populations
of lizards competing according to very similar rules [\rSL].}
$$
A + B  \to 2A, \qquad B + C  \to 2B, \qquad C + A  \to 2C.
$$
We notice immediately that there is a kind of cyclicity inherent in 
this set of reactions.  Ruijgrok and Ruijgrok have analyzed a system
of such particles which are simultaneously interacting according to
$$
A + 2B \to 3B, \qquad B + 2C \to 3C, \qquad C + 2A \to 3A,
$$
with relative rate $\alpha$ [\Ruijgrok].

Since each of the reactions conserves particle number we can normalize
to population densities $A+B+C = 1$.  Since $A$, $B$ and $C$ are all
positive, the configurations of the system can be represented by 
points in the interior of an equilateral triangle with side 
$2/\sqrt{3}$:  for any point inside this triangle the sum of the 
altitudes $A$, $B$ and $C$ to the three sides is 1.  The evolution of
the system is a flow on the triangle, described by the system of three
ODEs obtained from 
$$
\dot A = A(B-C) + \alpha A(AC-B^2)
$$
by cyclic permutations of $(ABC)$.

Notice that the center of the triangle $A=B=C=1/3$ is an equilibrium
point.  When $\alpha = 0$, the product $P := ABC$ is left invariant by
the flow:  all the solutions are periodic orbits along the closed
curves $P = {\rm const}$.  In general, 
$$
{1\over P(t)}{{\rm d}P(t)\over {\rm d}t} = -{9\over4}\alpha r^2(t),
$$
where $r(t)$ is the distance from the center of the triangle.  Thus
for $\alpha < 0$, the center $(1/3,1/3,1/3)$ is a stable equilibrium,
while for $\alpha > 0$ it is unstable; the $\alpha = 0$ model is not
structurally stable.

Although we have not introduced the formalism for finite topological 
spaces, this analysis motivates identification of the cyclicity of the 
two sets of reactions as topological nontriviality; certainly this
cyclicity can induce cycles in the evolution of the system.  Moreover,
in the sense that the system can be represented globally as a cycling
`ecology', or one at a stable or unstable equilibrium, rather than 
only in terms of individual, interacting particles, it is complex.

\medskip
\noindent{\bf 5.  Economics and politics}

\noindent Thinking of the preceding system as a model in population
biology leads us towards social science models of multiple interacting
agents.  By now we should expect cycles to occur in these 
systems---and they do, in very similar ways.  We begin by considering
an example, due to Scarf [\Scarf], which indicates difficulties with 
the general equilibrium model in economics [\Walras].

In this model there are multiple agents with initial endowments 
$w_i \in \R^k_{\ge0}$, where the components of $w_i$ represent amounts
of $k$ goods.  The prices of these goods are represented by another 
vector $p \in \R^k_{\ge0}$, so the initial wealth of agent $i$ is
$p\cdot w_i$.  By exchanging goods according to prices $p$, agent $i$
can afford any commodity bundle $x \in \R^k_{\ge0}$, provided 
$p\cdot x \le p\cdot w_i$.

The agents also have preferences among the different goods, modelled
by {\sl utility functions\/} $u_i : \R^k_{\ge0} \to \R$; 
$u_i(y) \ge u_i(x)$ means that agent $i$ prefers commodity bundle $y$ 
to bundle $x$.  We assume that for each agent $i$, 
$\{y \in \R^k_{\ge0} \mid u_i(y) \ge u_i(x)\}$ is a strictly convex
set and that $\nabla u_i \in \R^k_{>0}$.  At given prices $p$, agent
$i$ maximizes utility with a commodity bundle $x_i(p)$ satisfying
$\lambda p = \nabla u_i\bigl(x_i(p)\bigr)$, for some $\lambda > 0$;
this bundle is the agent's {\sl demand}.  Notice that any rescaling of
$p$ can be absorbed into $\lambda$; we follow tradition and rescale so
that $p\cdot p = k$, although we could equally well rescale $p$ so 
that $p\cdot (1,\ldots,1) = 1$; in either case we refer to the 
{\sl price simplex\/} of possible price vectors.  The {\sl excess 
demand\/} of agent $i$ at prices $p$ is $x_i(p) - w_i$; the 
{\sl aggregate excess demand\/} 
$v(p) := \sum_i\bigl(x_i(p) - w_i\bigr)$ is a continuous vector field
on the price simplex.

The aggregate excess demand defines a flow on the price simplex 
according to $\dot p = v(p)$.  That there is always an equilibrium 
point $v(p) = 0$ for such a flow is a consequence of Brouwer's fixed
point theorem [\Brouwer]; this is the {\sl topological\/} content 
[\Uzawa] of the Arrow-Debreu approach to the existence of competitive 
equilibria [\rAD].  Scarf demonstrated, however, that such equilibria 
can be globally unstable [\Scarf]:  Consider the utility function in a 
$k = 3$ good economy defined by $u_1(x) = \min (x_1,x_2)$.  Suppose 
this is the utility function for agent 1 and that there are two more 
agents with utility functions obtained from $u_1$ by cyclic 
permutations of $(x_1,x_2,x_3)$.  Let the initial endowments of the 
three agents be $(1,0,0)$ and its cyclic permutations, respectively.  
Then the first component of the aggregate excess demand is 
$$
v_1(p) = {-p_2\over p_1+p_2} + {p_3\over p_3+p_1};
$$
the other two components are obtained by cyclic permutation of
$(p_1,p_2,p_3)$.

The center point $p_1 = p_2 = p_3 = 1$ is an equilibrium point and, 
just as in the $\alpha = 0$ case of the example discussed in the
preceding section, the product $p_1p_2p_3$ is left invariant by the 
flow.  Thus the system has no stable equilibrium.  While the utility 
functions used in this example are convex, but not strictly convex, 
and the initial endowments are at an extreme, Scarf showed that there
is a family of such models (just as there was in Section~4) which have
strictly convex utility functions and inital endowments nonzero in 
each good, but which are also globally unstable [\Scarf].

In this economic model the discrete cycle of utility functions leads
to {\sl continuous\/} cyclic orbits just as the discrete cycle of 
reaction equations did in Ruijgrok and Ruijgrok's model.  But suppose
there are only a finite number of states for the system.  This is the
situation in political science models where each agent's utility
function is replaced with a {\sl preference order\/} among a finite 
set of alternatives:  a relation, denoted $\ge$, which is 
{\sl complete\/} (for all pairs of alternatives $a \ge b$ or 
$b \ge a$) and {\sl transitive\/} (if $a \ge b$ and $b \ge c$ then 
$a \ge c$) [\Arrow].  When $a \ge b$ and $b \ge a$, the agent with 
this preference order is {\sl indifferent\/} between $a$ and $b$; when 
only $a \ge b$, say, the agent {\sl strictly\/} prefers $a$ and we
write $a > b$.

The analogue of the aggregate excess demand is a map $f$ from 
preference {\sl profiles\/} (lists of the agents' preference orders) 
$p$ to directed graphs $f_p$.  A directed edge $a \leftarrow b$ in 
$f_p$ indicates that for profile $p$ the map $f$ chooses alternative 
$a$ over alternative $b$.  We call $f$ a {\sl voting rule\/} if for 
all profiles $p$, $f_p$ is {\sl complete\/} (for all pairs of 
alternatives $a \leftarrow b$ or $b \leftarrow a$) and Pareto (if
$a \ge b$ in each preference order in $p$ then $a \leftarrow b$ in
$f_p$).

More than 200 years ago Condorcet recognized that there are potential
problems with voting rules, namely that aggregation might produce
cycles rather than a definitive outcome [\Condorcet].  For example, 
suppose that there are three alternatives $\{a,b,c\}$ and three agents 
rank them in the orders $a > b > c$, $b > c > a$, and $c > a > b$.  
Given a choice between $b$ and $a$, a 2:1 majority prefers $a$; if 
they are offered the opportunity to switch from $a$ to $c$, again a 
majority will vote to do so; finally a majority also prefers $b$ to 
$c$, completing a cycle.  This cycle exists in the directed graph 
$f_p$ corresponding to the majority voting rule; it is the discrete 
analogue of the continuous cyclic orbits we have seen in the two 
preceding systems.

\medskip
\noindent{\bf 6.  Complexity and chaos}

\noindent These social science models, therefore, have features in 
common with the natural science models we considered in Sections~2, 3 
and 4.  At the more global scale defined by the aggregation 
mechanisms, the set of agents may be replaced by a cyclic (sub)market 
and a cyclic decision process, respectively.  Of course, if the 
utility functions of the economic agents are such that the aggregate 
excess demand is representable as the excess demand for a single agent 
then the market has a unique stable equilibrium.  Similarly, given a 
voting rule, the preferences of the agents may cohere to the extent 
that there is a definitive outcome; even more simply, the voting rule 
might be {\sl dictatorial}, \ie, dependent only on the preferences of 
a single, specified agent.  In all of these cases the model at the 
global scale is similar to the individual agent model; the topology is 
trivial and the systems are simple.

We emphasize, however, that complexity is inherent in social 
aggregation:  The theorems of Sonnenschein, Mantel, and Debreu show 
that for two or more goods and at least as many agents, the 
aggregate excess demand can be {\sl any\/} continuous vector field on 
the price simplex [\rSMD].  In particular, the flow need not have a
stable equilibrium:  As we have seen already in two dimensions, even
though the Poincar\'e-Bendixson theorem precludes more complicated 
behaviour [\rPB], the limit set may include cycles.  If there is any
time dependence from external effects, and in higher dimensional
markets, the Kupka-Smale theorem [\rKS] implies that {\sl chaotic\/} 
dynamics is structurally stable; systems with a unique or stable 
equilibrium point are far from generic.

For completely discrete voting models, Arrow's theorem imples that for 
more than two agents, under a reasonable condition%
\sfootnote*{This is the {\sl independence of irrelevant alternatives}, 
namely that the relation between $a$ and $b$ in $f_p$ depend only on 
the relations of $a$ and $b$ in the preference orders in $p$.}
on voting rules, any which are not dictatorial must contain cycles
[\Arrow].  We recently showed that in the latter situation, not only 
is the system complex, but it is also chaotic in the mathematical 
sense [\rMB]:  The {\sl topological entropy\/} [\rSPAKM], defined to 
be
$$
\lim_{n\to\infty} 
 {1\over n}\log(\hbox{number of $n$-periodic orbits in $f_p$})
$$
is positive exactly when there is a cycle in $f_p$; beyond identifying
the existence of chaos, it quantifies `how chaotic' the system is.  
Averaged over the space of voter preferences, the topological entropy 
measures the complexity of a voting rule, and averaged over voting 
rules, it measures the (lack of) coherence among voter preferences 
[\rMB].

\medskip
\noindent{\bf 7.  Consequences}

\noindent Aggregation rules define the way models of social systems
scale towards the global.  The larger scale model may be similar to 
the smaller scale one, as is the case for a dictatorial voting rule, 
for example.  In equivalently simple physical models, rescaling simply 
renormalizes the state variables and their interactions, but 
introduces no new ones.  Complexity emerges when nontrivial topology 
exists at the more global scale.  This is a cycle in each of our 
examples, and it induces qualitatively different models in both 
natural and social systems.

The attendant chaos has profound consequences for simulations of these
systems:  Fine grained prediction is impossible---only certain 
statistical properties of the evolution are robust.  This fact is
appreciated heuristically in some of the economics and sociology 
literature [\rSB], but seems to receive less emphasis in discussions 
of specific multiagent simulation platforms [\rMBLAEA].  It will, 
nevertheless, determine to which problems such programs can be applied 
successfully.

Systems of multiple software agents, interacting to effect some 
(un)intentional distributed computation, are subject to exactly the
same analysis.  Whether their interactions are decision theoretic 
[\rZRER] or market oriented [\rWCW], designers and users must be aware 
of the possibilities for complexity and chaos.  Even when the 
artificial agents are physical---autonomous robots [\rHT]---attempts
at designing coordinated action [\rBWM] should be informed by these
same considerations.

Particularly in the context of deploying interacting software or 
hardware agents for some specific task, but also in the context of
modelling or simulating multiagent social systems, success may depend
on controlling the degree of chaos in the resulting complex system.
As we noted at the end of the last section, the topological entropy,
for example, measures how chaotic a system is and allows us to 
identify the sources of complexity, but this alone is insufficient for 
control.  There are several possibilities:  We have primarily 
considered immutable agents; there are also models with adaptive 
agents [\rHM] which one might hope to have better generic behaviour.  
The population biology example of Section~4, however, models a 
particular kind of adaptation, so chaos is probably no less generic in
complex adaptive systems.  Agents with `higher' rationality have also 
been modelled [\rGDT]; these are agents which make internal models for 
the behaviour of the other agents with which they interact.  This 
approach, while possibly more realistic for modelling human agents,
seems to flirt with self-reference and undecidability [\Godel], and 
must, at best, constrain the depth of the internal models to achieve 
acceptable computational efficiency [\rVD].  Most generally, it seems 
that models for complex social systems should include some form of
`back reaction' from the more global aggregation scale to the local
individual agent scale.  Such a back reaction might affect the 
agents' preferences or the way they interact, but must be carefully
implemented to model interesting complex systems which balance
precariously between simplicity and extreme chaos.

\medskip
\noindent{\bf Acknowledgements}
\nobreak

\nobreak
\noindent This paper reports on part of a collaborative project with 
Thad Brown, Gary Doolen, Brosl Hasslacher, Ronnie Mainieri and Mark 
Tilden.  I also thank Mike Freedman, Melanie Quong, Leslie Smith and 
Peter Teichner for discussions on various aspects of this work, and 
Bruce Boghosian for inviting me to speak in the modelling and
simulation session at the ICCS.  I gratefully acknowledge support from
the John Deere Foundation and from DISA.

\medskip
\global\setbox1=\hbox{[00]\enspace}
\parindent=\wd1

\noindent{\bf References}

\item{[\Segel]}
L. A. Segel,
``The theoretician grapples with complex systems'',
SFI working paper 93-05-032 (1993).

\parskip=0pt
\item{[\rMB]}
\dajm\ and T. A. Brown,
``Statistical mechanics of voting''.
CSC/IPS/UCSD/ UM preprint (1997).

\item{[\rBH]}
J. Barnes and P. Hut,
``A hierarchical $O(N\log N)$ force-calculation algorithm'',
\Na\ {\bf 324} (1986) 446--449.

\item{[\Sorkin]}
R. D. Sorkin,
``Introduction to topological geons'',
in P. G. Bergmann and V. de Sabbata, eds.,
{\sl Topological Properties and Global Structure of Space-Time},
proceedings of the NATO Advanced Study Institute, Erice, Italy,
  12--22 May 1985
(New York:  Plenum 1986) 249--270.

\item{[\Kelvin]}
W. H. Thomson, Baron Kelvin,
``On vortex atoms'',
\PRSE\ {\bf VI} (1867) 94--105.

\item{[\Helmholtz]}
H. L. F. von Helmholtz,
``{\it \"Uber Integrale der hydrodynamischen Gleichungen, welche den
  Wir\-belbewegung entsprechen}'',
\JRAM\ {\bf LV} (1858) 25--55.

\item{[\Leonard]}
A. Leonard,
``Vortex methods for flow simulation'',
\JCP\ {\bf 37} (1980) 289--335.

\item{[\Ruijgrok]}
Th.\ Ruijgrok and M. Ruijgrok,
``A reaction-diffusion equation for a cyclic system with three
  components'',
\JSP\ {\bf 87} (1997) 1145--1164.

\item{[\rSMD]}
H. Sonnenschein,
``Do Walras' identity and continuity characterize the class of 
  community excess demand functions?'',
\JET\ {\bf 6} (1973) 345--354;\hfb
R. R. Mantel,
``On the characterization of aggregate excess demand'',
\JET\ {\bf 7} (1974) 348--353;\hfb
G. Debreu,
``Excess demand functions'',
\JME\ {\bf 1} (1974) 15--21.

\item{[\Arrow]}
K. J. Arrow,
{\sl Social Choice and Individual Values\/}
(New York:  Wiley 1951).

\item{[\rBP]}
J. P. Boris,
``New directions in computational fluid dynamics'',
\ARFM\ {\bf 21} (1989) 345--385;\hfb
R. Peyret, ed.,
{\sl Handbook of Computational Fluid Mechanics\/}
(San Diego:  Academic Press 1996).

\item{[\Rosenhead]}
L. Rosenhead,
``The formation of vortices from a surface of discontinuity'',
\PRSLA\ {\bf 134} (1931) 170--192.

\item{[\Aref]}
H. Aref,
``Integrable, chaotic, and turbulent vortex motion in two-dimensional
  flows'',
\ARFM\ {\bf 15} (1983) 345--389.

\item{[\Christiansen]}
J. P. Christiansen,
``Numerical simulation of hydrodynamics by the method of point 
  vortices'',
\JCP\ {\bf 13} (1973) 363--379.

\item{[\Chorin]}
A. J. Chorin, 
``Numerical study of slightly viscous flow'',
\JFM\ {\bf 57} (1973) 785--796.

\item{[\Hald]}
O. H. Hald,
``Convergence of vortex methods for Euler's equations.  II.'',
\SIAMJNA\ {\bf 16} (1979) 726--755.

\item{[\rSL]}
B. Sinervo and C. M. Lively,
``The rock-paper-scissors game and the evolution of alternative male
  strategies'',
\Na\ {\bf 380} (1996) 240--243.

\item{[\Scarf]}
H. Scarf,
``Some examples of global instability of the competitive 
  equilibrium'',
\IER\ {\bf 1} (1960) 157--172.

\item{[\Walras]}
L. Walras,
{\it El\`ements d'\'economie politique pure, ou, Th\'eorie de la 
  richesse sociale\/}
(Lausanne:  Corbaz 1874--7).

\item{[\Brouwer]}
L. E. J. Brouwer,
``{\it \"Uber Abbildung von Mannigfaltigkeiten}'',
\MA\ {\bf 71} (1912) 97--115.

\item{[\Uzawa]}
H. Uzawa,
``Walras' existence theorem and Brouwer's fixed point theorem'',
\ESQ\ {\bf 12} (1962) 59--62.

\item{[\rAD]}
K. J. Arrow and G. Debreu,
``Existence of an equilibrium for a competitive economy'',
\E\ {\bf 22} (1954) 265--290.

\item{[\Condorcet]}
M.-J.-A.-N. de Caritat, Marquis de Condorcet,
{\it Essai sur l'application de l'analyse \`a la probabilit\'e des 
     d\'ecisions rendues \`a la pluralit\'e des voix\/}
(Paris:  {\it l'Imprim\`erie Royale\/} 1785).

\item{[\rPB]}
J. H. Poincar\'e,
``{\it M\`emoire sur les courbes d\'efinies par les \'equations
       diff\'erentielles, I.}'',
\JMPA, {\sl 3.\ s\'erie\/} {\bf 7} (1881) 375--422; 
{\it II.} {\bf 8} (1882) 251--286; 
{\it III.} {\sl 4.\ s\'erie\/} {\bf 1} (1885) 167--244;
{\it IV.} {\bf 2} (1886) 151--217;\hfb
I. Bendixson,
``{\it Sur les courbes d\'efinies par les \'equations 
       diff\'erentielles\/}'',
\AcM\ {\bf 24} (1901) 1--88.

\item{[\rKS]}
I. Kupka,
``{\it Contributions \`a la th\'eorie des champs g\'en\'erique\/}'',
\CDE\ {\bf 2} (1963) 457--484;\hfb
S. Smale,
``Stable manifolds for differential equations and diffeomorphisms'',
\ANASP, {\sl ser.\ III\/} {\bf 17} (1963) 97--116.

\item{[\rSPAKM]}
C. E. Shannon,
``A mathematical theory of communication'',
\BSTJ\ {\bf 27} (1948) 379--423; 623--656;\hfb
W. Parry,
``Intrinsic Markov chains'',
\TAMS\ {\bf 112} (1964) 55--66;\hfb
R. L. Adler, A. G. Konheim and M. H. McAndrew,
``Topological entropy'',
\TAMS\ {\bf 114} (1965) 309--319.

\item{[\rSB]}
A. C. S\'eror,
``Simulation of complex organizational processes:  a review of 
  methods and their epistemological foundations'',
in N. Gilbert and J. Doran, eds.,
{\sl Simulating Societies:  The Computer Simulation of Social 
  Phenomena\/}
(London:  UCL Press 1994) 19--40;\hfb
D. Byrne,
``Simulation---a way forward?'',
\SRO\ {\bf 2} (1997)\break
http://www.socresonline.org.uk/socresonline/2/2/4.html.

\item{[\rMBLAEA]}
N. Minar, R. Burkhart, C. Langton and M. Askenazi,
``The Swarm simulation system:  a toolkit for building multi-agent
  simulations'',
SFI working paper 96-06-042 (1996);\hfb
J. M. Epstein and R. Axtell,
{\sl Artificial Societies:  Social Science from the Bottom Up\/}
(Cambridge, MA:  MIT Press 1996).

\item{[\rZRER]}
G. Zlotkin and J. S. Rosenschein,
``Mechanism design for automated negotiation, and its application
  to task oriented domains'',
\AI\ {\bf 86} (1996) 195--244;\hfb
E. Ephrati and J. S. Rosenschein,
``Deriving consensus in multiagent systems'',
\AI\ {\bf 87} (1996) 21--74.

\item{[\rWCW]}
M. P. Wellman,
``The economic approach to artificial intelligence'',
\ACMCS\ {\bf 27} (1995) 360--262;\hfb
J. Q. Cheng and M. P. Wellman,
``The WALRAS algorithm:  a convergent distributed implementation of
  general equilibrium outcomes'',
to appear in \CE

\item{[\rHT]}
\brosl\ and M. W. Tilden,
``Living machines'',
\RAS\ {\bf 15} (1995) 143--169.

\item{[\rBWM]}
G. Beni and J. Wang,
``Theoretical problems for the realization of distributed robotic
  systems'',
in {\sl Proceedings of the 1991 IEEE International Conference on 
  Robotics and Automation}, 
Sacramento, CA 9--11 April 1991
(Los Alamitos, CA:  IEEE Computer Society Press 1991) 1914--1920;\hfb
M. J. Mataric,
``Distributed approaches to behavior control'',
in P. S. Schenker, ed.,
{\sl Sensor Fusion V}, 
proceedings of the SPIE conference, {\bf 1828}, 
Boston, MA, 15--17 November 1992
(Bellingham, WA:  SPIE 1992) 373--382.

\item{[\rHM]}
P. T. Hraber and B. T. Milne,
``Community assembly in a model ecosystem'',
SFI working paper 96-12-094 (1996).

\item{[\rGDT]}
P. J. Gmytrasiewicz and E. H. Durfee,
``A rigorous, operational formalization of recursive modelling'',
in {\sl ICMAS-95}, Proceedings of the First International 
  Conference on Multi-Agent Systems, 
San Francisco, 12--14 June 1995
(Menlo Park, CA:  AAAI Press 1995) 125--132;\hfb
L. Tesfatsion,
``A trade network game with endogenous partner selection'',
in H. M. Amman, B. Rustem and A. B. Whinston, eds.,
{\sl Computational Approaches to Economic Problems\/}
(Boston:  Kluwer Academic 1997) 249--269.

\item{[\Godel]}
K. G\"odel,
``{\it \"Uber formal unentscheidbare S\"atze der {\sl Principia
   Mathematica\/} und verwandter System I.}'',
\MMP\ {\bf 38} (1931) 173--198.

\item{[\rVD]}
J. M. Vidal and E. H. Durfee,
``Using recursive models effectively'',
in M. Wooldridge, J. P. M\"uller and M. Tambe, eds.,
{\sl Intelligent Agents II:  Agent Theories, Architectures, and
  Languages}, proceedings of the IJACI'95 Workshop (ATAL), 
Montr\'eal, 19--20 August 1995, LNCS {\bf 1037}
(New York:  Springer-Verlag 1996) 171--186.

\bye